\documentclass[graybox]{svmult}

\usepackage{mathptmx}       
\usepackage{helvet}         
\usepackage{courier}        
\usepackage{type1cm}        
\usepackage{makeidx}         
\usepackage{graphicx}        
\usepackage{multicol}        
\usepackage[bottom]{footmisc}

\makeindex             

\begin{document}

\title*{RR Lyrae type stars, ST Boo and RR Leo: 2007 Observations and the preliminary results of
the frequency analaysis}
\titlerunning{ST Boo and RR Leo: Observations and frequency analysis} 
\author{Fehmi EKMEK\c{C}\.{I}, Lale \c{C}EL\.{I}K, H. Volkan \c{S}ENAVCI}
\authorrunning{Ekmek\c{c}i, \c{C}elik \& \c{S}enavc{\i}} 
\institute{Fehmi Ekmek\c{c}i, Lale \c{C}elik, H. Volkan \c{S}enavc{\i} \at Ankara University, Faculty of Science, Department of Astronomy and Space Sciences, 06100, Tando\u{g}an, Ankara, Turkey, \email{fekmekci@science.ankara.edu.tr}}

%
%
\maketitle


\abstract{We present BVR light curves of pulsating stars, ST Boo and RR Leo, obtained between March and September 2007 at the Ankara University Observatory (AUG) and the T\"{U}B{\.I}TAK National Observatory (TUG). Although these observational data are insufficient to obtain the reliable results for a frequency analysis of ST Boo and RR Leo stars, in this study, we tried to investigate the pulsation phenomena of these two stars, as an overview, using the Period04 software package. As preliminary results, we present the possible frequencies for ST Boo and RR Leo.}

\section{Observations and Results}
\label{sec:1}

CCD observations of ST Boo and RR Leo were carried out by using an Apogee ALTA $U47+CCD$ camera ($1024\times1024$ pixels) with BVR filters mounted on both 40 cm Schmidt-Cassegrain telescopes of the Ankara University Observatory (AUG) and the T\"{U}B{\.I}TAK National Observatory (TUG) between March and September 2007. BVR light curves of both ST Boo and RR Leo were normalized to maximum light level to construct the data set for simultaneous multiple-frequency analysis using Period04 (V 1.0) (\cite{Lenz05}) which has a Fourier analysis definition of

\begin{equation}
 f(t) = Z + \sum_{i} A_{i} sin(2\pi(\Omega_{i} t + \phi_{i})).   
\end{equation}

\begin{figure}
\vspace{3.5cm}
\includegraphics{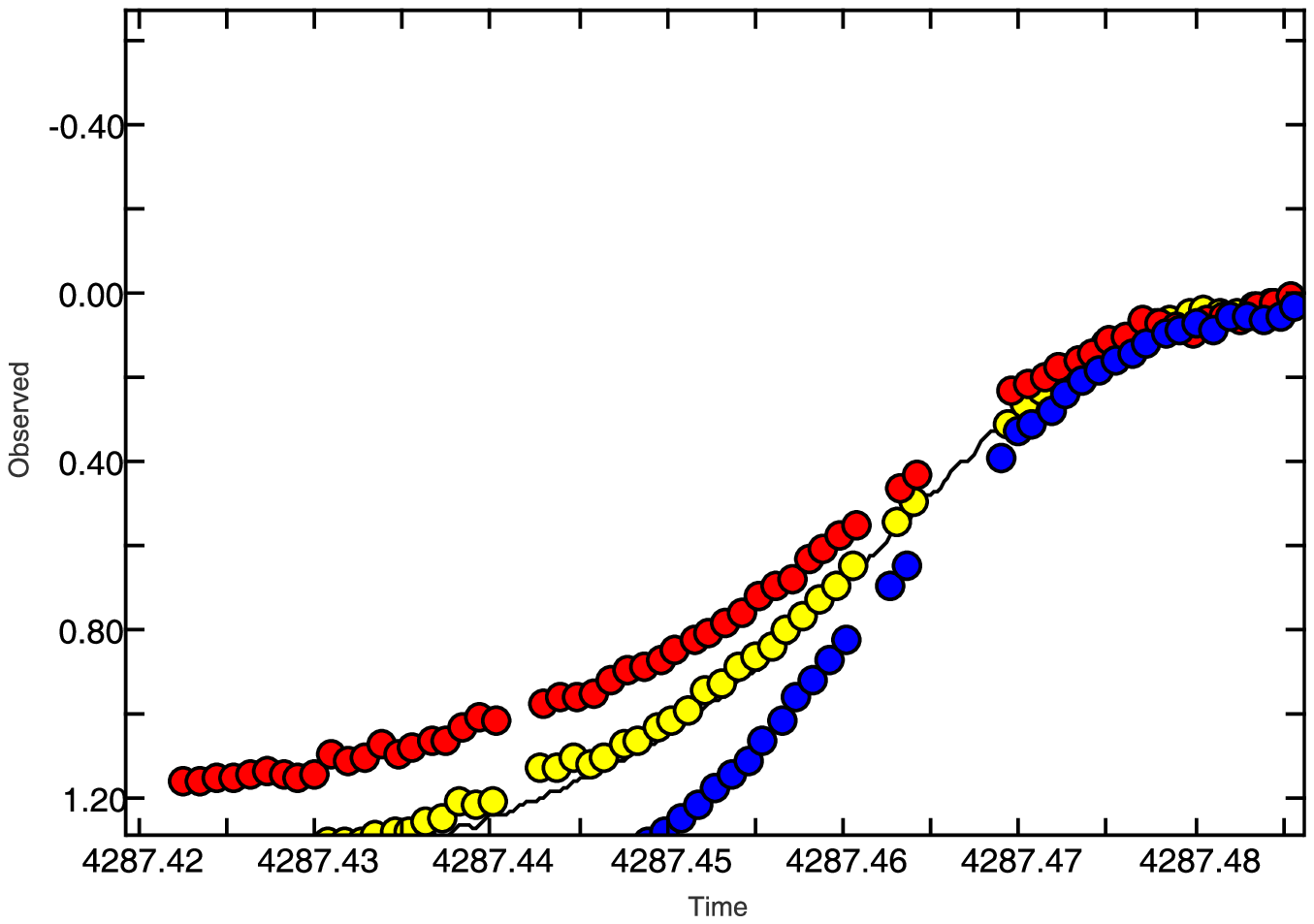}
\includegraphics{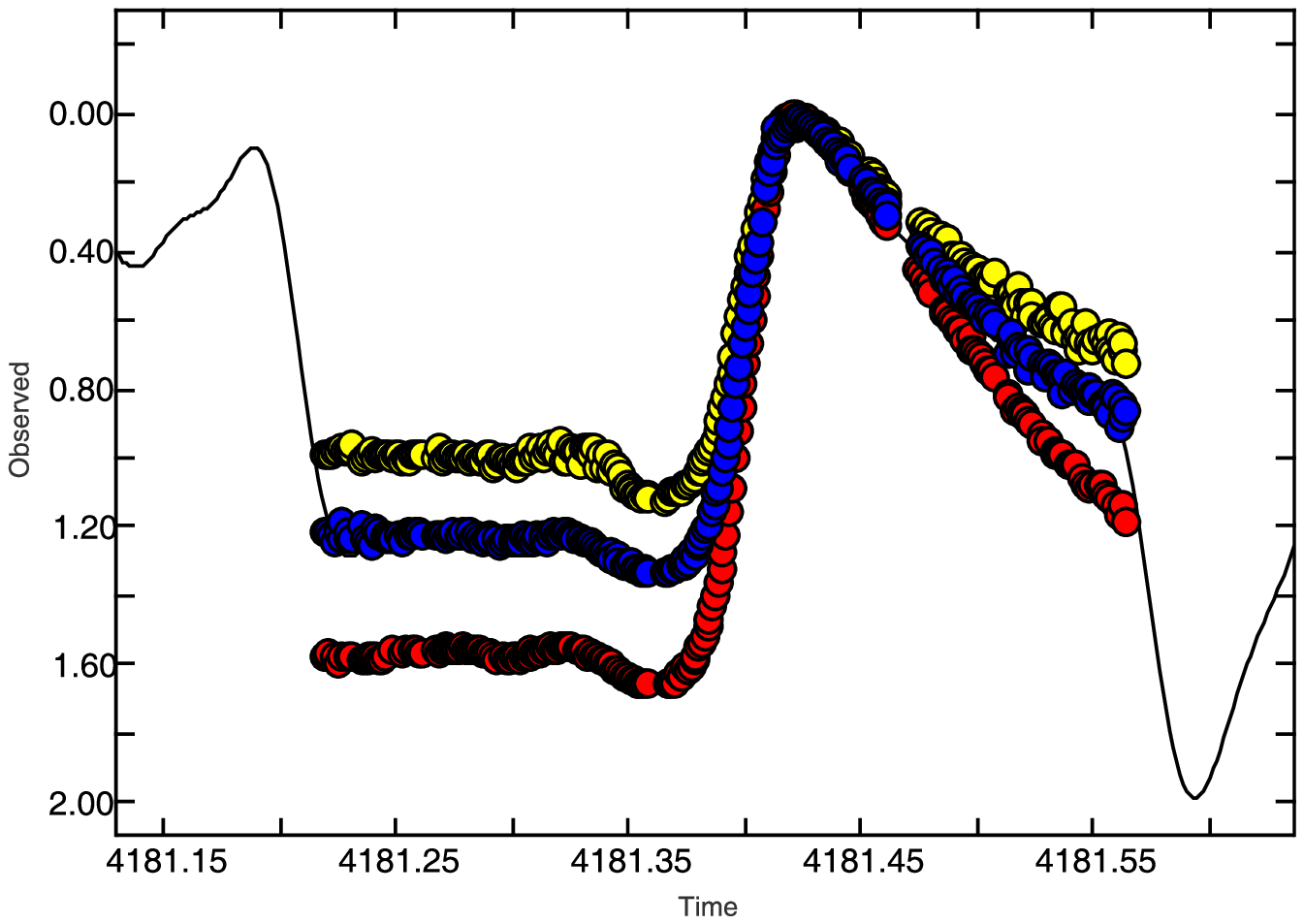}
\caption{An example of the light curve measurements of ST Boo (left panel) and RR Leo (right panel) together with the fit curve of multi-frequency solution. The axis of the time is in unit of HJD(2450000+...) and observed BVR is in normalized values to
maximum level of the light curve}
\label{fig:1}
\end{figure}

The results of multi-frequency solutions, with their errors calculated based on Monte Carlo Simulation, for ST Boo and RR Leo are given in Table~\ref{tab:1}. Fig.~\ref{fig:1} shows some of the light curve data with the fit curve of multi-frequency solutions for ST Boo and RR Leo. Clearly, it must be included more and more photometric data in the frequency analysis to have more definite and reliable results for both of these pulsating stars. 
         
\begin{table}
\tiny
\caption{The results of multiple-frequency analysis of ST Boo and RR Leo}
\label{tab:1}     
\begin{tabular}{p{1.5cm}p{1.5cm}p{1.3cm}p{1.5cm}p{1.5cm}p{1.3cm}}
\hline\noalign{\smallskip}
             &   ST Boo     &     &           &       &     \\
\hline\noalign{\smallskip}                    
f($cd^{-1}$) & Amp.(mag.) & S/N & f($cd^{-1}$) & Amp.(mag.) & S/N \\
\noalign{\smallskip}\svhline\noalign{\smallskip}
$4.201\pm0.079$& $0.553\pm0.664$ & 2.84 & $18.925\pm0.663$ & $0.012\pm0.022$ & 32.02 \\ 
$4.221\pm0.668$& $0.171\pm0.754$ & 617.14 & $28.315\pm0.845$& $0.011\pm0.012$ & 19.93 \\
$0.187\pm0.005$& $0.127\pm0.209$ & 455.01 & $16.236\pm0.342$ & $0.010\pm0.033$& 19.81 \\ 
$7.311\pm0.034$& $0.119\pm0.072$ & 538.73 & $20.333\pm0.695$ & $0.008\pm0.026$ & 25.20 \\
$6.337\pm0.027$& $0.097\pm0.071$ & 419.63 & $33.717\pm0.370$ & $0.008\pm0.004$ & 9.53 \\
$10.247\pm2.389$& $0.088\pm0.143$& 398.35 & $28.724\pm0.176$ & $0.007\pm0.009$ & 12.22 \\
$10.649\pm0.504$& $0.085\pm0.178$& 365.89 & $23.532\pm0.418$ & $0.006\pm0.014$ & 10.29 \\
$7.500\pm0.082$& $0.076\pm0.114$ & 348.95 & $602.962\pm0.856$ & $0.005\pm0.002$ & 5.96 \\ 
$17.073\pm0.093$& $0.043\pm0.055$ & 87.02 & $35.863\pm0.108$ & $0.005\pm0.006$ & 5.00 \\                                             
$0.227\pm0.060$& $0.037\pm0.560$ & 133.81 & $604.857\pm0.239$& $0.004\pm0.002$ & 5.22\\ 
$10.875\pm1.406$& $0.033\pm0.139$& 138.95 & $510.716\pm0.157$& $0.004\pm0.002$ & 5.53\\ 
$3.113\pm0.540$ & $0.028\pm0.099$ & 96.61 & $30.484\pm1.174$& $0.003\pm0.006$ & 6.18 \\ 
$21.491\pm0.105$& $0.024\pm0.020$ & 65.53 & $568.959\pm7.439$& $0.003\pm0.002$ & 4.91\\
$11.980\pm0.313$ & $0.024\pm0.043$ & 83.07 & $599.575\pm0.112$& $0.003\pm0.003$ & 4.21\\
$14.403\pm0.093$ & $0.024\pm0.043$ & 49.74 & $566.738\pm0.198$& $0.003\pm0.002$ & 5.00\\
$16.399\pm3.804$ & $0.024\pm0.055$ & 45.08 & $559.057\pm0.150$& $0.003\pm0.002$ & 4.03\\
$6.465\pm0.255$ & $0.021\pm0.135$ & 90.25 &  $47.909\pm0.366$& $0.003\pm0.002$ & 4.28\\
$24.493\pm2.307$ & $0.016\pm0.020$ & 23.59 & $562.309\pm7.929$& $0.002\pm0.002$ & 4.29\\
$21.501\pm0.250$ & $0.012\pm0.016$ & 31.06 &  - & - & - \\
\noalign{\smallskip}\hline\noalign{\smallskip}
             &   RR Leo    &     &         &        &         \\
\hline\noalign{\smallskip}                    
f($cd^{-1}$) & Amp.(mag.) & S/N & f($cd^{-1}$) & Amp.(mag.) & S/N \\
\noalign{\smallskip}\svhline\noalign{\smallskip}
$3.418\pm0.001$& $0.672\pm0.041$ & 2.88 & $10.788\pm0.003$ & $0.055\pm0.016$ & 10.81 \\
$8.634\pm0.001$& $0.229\pm0.026$ & 35.15 & $19.478\pm0.001$ & $0.049\pm0.010$ & 14.42 \\ 
$4.541\pm0.206$& $0.166\pm0.060$ & 17.06 & $16.500\pm0.239$ & $0.031\pm0.017$ & 7.18 \\
$13.469\pm0.001$& $0.132\pm0.022$& 28.57 & $21.518\pm0.037$ & $0.031\pm0.006$ & 9.97 \\
$6.962\pm0.057$ & $0.075\pm0.038$ & 10.08 & $24.141\pm0.001$ & $0.030\pm0.005$ & 9.44 \\
$2.259\pm0.246$ & $0.062\pm0.049$ & 4.04 & $27.494\pm0.159$ & $0.016\pm0.004$ & 4.48 \\
\noalign{\smallskip}\hline\noalign{\smallskip}             
\end{tabular}
\end{table}

\end{document}